\documentclass[11pt,a4,onecolumn,secnumarabic,amssymb, amsmath, nofootinbib,tightenlines,
nobibnotes, aps, prl,epsfig]{revtex4}
\usepackage{graphicx}% Include figure files
\usepackage{dcolumn}% Align table columns on decimal point
\usepackage{bm}% bold math

\begin{document}
\preprint{APS/123-QED}
\title{Decoupling of the DGLAP evolution equations by Laplace method }% Force line breaks with \\

\author{G.R. Boroun}%
\email{boroun@razi.ac.ir; grboroun@gmail.com}
 \author{S. Zarrin}%
\email{salah.zarin@gmail.com }
\author{F. Teimoury}
\affiliation{ Physics Department, Razi University, Kermanshah
67149, Iran}% \textbackslash\textbackslash
\date{\today}% It is always \today, today,

\begin{abstract}
In this paper, we derive two second- order of differential
equation for the gluon  and singlet distribution functions by
using the Laplace transform method. We decoupled the solutions of
the singlet and gluon distributions
 into the initial conditions (function and derivative of the function) at the
virtuality $Q_{0}^{2}$ separately as these solutions are defined
by:
\begin{eqnarray}
  F_{2}^{s}(x,Q^{2})&=&\mathcal{F}(F_{s0}, \partial F_{s0}
)\nonumber\\
 &&\mathrm{and}\nonumber\\
G(x,Q^{2})&=&\mathcal{G}(G_{0}, \partial G_{0}).\nonumber
\end{eqnarray}
 We compared our results with the MSTW parameterization
and the experimental measurements of $F_{2}^{p}(x,Q^{2})$.
\end{abstract}
%\pacs{11.55Jy, 12.38.-t, 14.70.Dj}%PACS, the Physics and Astronomy
                              %Classification Scheme.
%\keywords{Gluon and singlet exponents; DGLAP evolution equations;
%Hard Pomeron; Small-$x$} %Use showkeys class option if keyword
\maketitle
%%%%%%%%%%%%%%%%%%%%%%%%%%%%%%%%%%%%%%%%%%%%%%%%%%%%%%%%%%%%%%%%%
\section{Introduction}
The Dokshitzer-Gribov-Lipatov-Altarelli-Parisi (DGLAP) [1-3]
evolution equations are a set of the integro-differential
equations as can be evaluated the parton distribution functions
(PDFs) into the $Q^{2}$ evolution (where $Q ^{2}$ is the four
momenta transfer in a deep inelastic scattering (DIS) process).
The DIS structure functions provide our information about the deep
structure of the hadrons and parton densities at collider and
coefficient functions which contain information about the
boson-parton interaction. These evolution equations are
fundamental tools to test perturbative quantum chromodynamics
(pQCD)  in DIS experiments where the density functions can be
evaluated by solving the DGLAP
equations into the initial distributions.\\
The structure function $F _{2} ( x , Q ^{2} )$ reflects the
momentum distributions of the partons in the nucleon. But the
gluon distribution $G( x , Q ^{2} )$ does not appear in the
experimental directly, and it is determined only through the quark
distributions in conjunction with the DGLAP evolution equations.
It is also important to know the gluon distribution inside a
hadron at low $x$ because gluons are expected to be dominate in
this region. Traditionally, gluon and quark distribution functions
have been determined using the two coupled integral-differential
(DGLAP) equations. The solutions of the unpolarized DGLAP
evolution equations decoupled have been discussed considerably
over the last years [4-7]. There exist two main classes of
approaches: those that solve the equation indirectly in $s$-space
at the Laplace transform method [4-6], and those  that decouple
the DGLAP equations directly in x-space with define a hard pomeron
behavior for the distribution functions [7]. In Refs.[4-6],
authors have derived the gluon distribution and the singlet
structure function into the initial conditions
$G_{0}(x,Q_{0}^{2})$ and $F_{s0}(x,Q_{0}^{2})$. The decoupled
solutions for the gluon distribution and the singlet structure
function determined by
\begin{eqnarray}
F_{2}^{s}(x,Q^{2})=\mathcal{F}(F_{s0}, G_{0})~~~ \mathrm{and}
~~~~G(x,Q^{2})=\mathcal{G}(F_{s0}, G_{0}).,
\end{eqnarray}
where the functions $\mathcal{F}$ and $\mathcal{G}$ are determined
by the splitting functions into the both initial conditions.
However this method is not general, because solutions of the
coupled DGLAP equations are possible under certain conditions. In
Ref.[7], a set of formulae to extract two second-order independent
differential equation for the gluon and singlet distribution
functions have derived. In this method, the singlet quark and
gluon distributions have the same high-energy behavior at low-$x$
according to the hard- pomeron behavior for parton distributions.
Decoupling of the  DGLAP evolution equations, for the functions
$F_{2}^{s}(x,Q^{2})$ and $G(x,Q^{2})$, determined as a function of
their initial parameterisation at the starting scale $Q_{0}^{2}$
as
\begin{eqnarray}
F_{2}^{s}(x,Q^{2})=\mathcal{F}(\partial F_{s0}, F_{s0})~~~
\mathrm{and} ~~~~G(x,Q^{2})=\mathcal{G}(\partial G_{0}, G_{0}),
\end{eqnarray}
but this method is dependence to the model behavior for the parton
distribution functions.\\
Now we would like to find an explicit and general solution for the
 decoupling of the DGLAP evolution equations. This method is
independent of the parton distribution behavior. In this paper,
the
 DGLAP evolution equations coupled convert to the two second order
differential equation which define $G(x,Q^{2})$ and
$F_{2}^{s}(x,Q^{2})$ in terms of the individual initial conditions
$\{G(x,Q_{0}^{2}), {\partial}G(x,Q_{0}^{2})\}~~ \mathrm{and}~~
\{F_{2}^{s}(x,Q_{0}^{2}), {\partial}F_{2}^{s}(x,Q_{0}^{2})\}$
respectively.\\
%%%%%%%%%%%%%%%%%%%%%%%%%%%%%%%%%%%%%%%%%%%%%%%%%%%%%%%%%%%%%%
\section{Method}
The DGLAP evolution equations can be written by
$$
\frac{4\pi}{\alpha_{s}(Q^{2})}\frac{\partial
F_{2}^{s}(x,Q^{2})}{\partial
\ln(Q^{2})}=\left[4+\frac{16}{3}\ln(\frac{1-x}{x})\right]F_{2}^{s}(x,Q^{2})+\frac{16x}{3}\int_{x}^{1}\left(\frac{F_{2}^{s}(z,Q^{2})}{z}-\frac{F_{2}^{s}(x,Q^2)}{x}\right)
\frac{dz}{z-x}
$$
\begin{equation}
-\frac{8}{3}x\int_{x}^{1}F_{2}^{s}(z,Q^2)\left(1+\frac{x}{z}\right)\frac{dz}{z^{2}}+2n_{f}x\int_{x}^{1}G(z,Q^{2})\left(1-2\frac{x}{z}+2\frac{x^{2}}{z^2}\right)\frac{dz}{z^{2}},
\end{equation}
and
$$
\frac{4\pi}{\alpha_{s}(Q^{2})}\frac{\partial G(x,Q^{2})}{\partial
\ln(Q^{2})}=\left[\frac{33-2n_{f}}{3}+12\ln(\frac{1-x}{x})\right]G(x,Q^{2})+12x\int_{x}^{1}\left(\frac{G(z,Q^{2})}{z}-\frac{G(x,Q^{2})}{x}\right)\frac{dz}{z-x}
$$
\begin{equation}
+12x\int_{x}^{1}G(z,Q^{2})\left(\frac{z}{x}-2+\frac{x}{z}-\frac{x^{2}}{z^{2}}\right)\frac{dz}{z^{2}}+\frac{8}{3}\int_{x}^{1}F_{2}^{s}(z,Q^{2})\left(1+\left(1-\frac{x}{z}\right)^{2}\right)\frac{dz}{z}.
\end{equation}
These equations are coupled into the singlet and gluon structure
functions. Let us introduce the variables $ v\equiv \ln(1/x) ~~
\mathrm{and}~~ w\equiv \ln(1/z)$ and note that
$F_{2}^{s}(e^{-v},Q^{2})\equiv\hat {F}_{2}^{s}(v,Q^{2}) ~~
\mathrm{and}~~ G(e^{-v},Q^{2})\equiv\hat{G}(v,Q^{2})$. Using
Laplace transform method in $v$-space [6-6,8-10], we can define
the distribution functions in $s$-space as
\begin{equation}
f(s,Q^2)=\mathcal{L}\left[\hat{F}_{2}^{s}(v,Q^{2});s\right]=\int_{0}^{\infty}\hat{F}_{2}^{s}(v,Q^{2})e^{-sv}dv
\ \ ,\ \
g(s,Q^2)=\mathcal{L}\left[\hat{G}(v,Q^{2});s\right]=\int_{0}^{\infty}\hat{G}_{s}(v,Q^{2})e^{-sv}dv,
\end{equation}
and
\begin{equation}
\mathcal{L}\left[\frac{\partial \hat{F}_{2}^{s}}{\partial
w}(w,Q^{2});\right]=sf(s,Q^{2}) \ \ , \ \
\mathcal{L}\left[\frac{\partial \hat{G}_{s}}{\partial
w}(w,Q^{2});\right]=sg(s,Q^{2}).
\end{equation}
We know that $F_{2}^{s}(v=0,Q^{2})=G(v=0,Q^{2})=0 $ and  introduce
the new variable $\tau$ as $ \frac{d \tau(Q^{2})}{d
\log(Q^{2})}=\frac{\alpha_{s}(Q^{2})} {4\pi}$ for the $Q^{2}$
dependence. Therefore, the coupled first order differential
equations in $s$-space  can be rewritten:
$$
\frac{\partial
f}{\partial\tau}(s,\tau)=\Phi_{f}(s)f(s,\tau)+\Theta_{f}(s)g(s,\tau),
$$
and
\begin{equation}
\frac{\partial
g}{\partial\tau}(s,\tau)=\Phi_{g}(s)g(s,\tau)+\Theta_{g}(s)f(s,\tau),
\end{equation}
where the coefficient functions $\Theta(s)$ and  $\Phi(s)$ are [4] \\
\begin{equation}
\Phi_{f}\left(s\right)=4-\frac{8}{3}\left(\frac{1}{s+1}+\frac{1}{s+2}+2\left(\psi(s+1)+\gamma_{E}\right)\right),
\end{equation}
\begin{equation}
\Theta_{f}\left(s\right)=2n_{f}\left(\frac{1}{s+1}-\frac{2}{s+2}+\frac{2}{s+3}\right),
\end{equation}
\begin{equation}
\Phi_{g}\left(s\right)=\frac{33-2n_{f}}{3}+12\left(\frac{1}{s}-\frac{2}{s+1}+\frac{1}{s+2}-\frac{1}{s+3}-\psi(s+1)-\gamma_{E}\right),
\end{equation}
and
\begin{equation}
\Theta_{g}\left(s\right)=\frac{8}{3}\left(\frac{2}{s}-\frac{2}{s+1}+\frac{1}{s+2}\right).
\end{equation}
Here $\psi(s)$ is the digamma function, $\gamma_{E}= 0.5772156 . .
. $ is Euler$^{,}$s constant and $n_{f}$ is the number of quark
flavors. Now let us find our solution for decoupling of the
functions $f$ and $g$  from the above equations. The explicit form
of the  two second order differential equation in $s$-space are
given by
\begin{equation}
\frac{\partial^{2}\digamma}{\partial^{2}\tau}(s,\tau)-\left(\Phi_{f}(s)+\Phi_{g}(s)\right)\frac{\partial
\digamma}{\partial
\tau}(s,\tau)-\left(\Theta_{f}(s)\Theta_{g}(s)-\Phi_{f}(s)\Phi_{g}(s)\right)\digamma(s,\tau)=0,~~(\digamma=f~~
\mathrm{or}~~g).
\end{equation}
Solving these equations, we obtain two independent solution as
$$
f(s,\tau)=c_{1}\exp\left(\frac{1}{2}(R_{1}(s)+R(s))\tau\right)+c_{2}\exp\left(\frac{1}{2}(R_{1}(s)-R(s))\tau\right),
$$
and
\begin{equation}
g(s,\tau)=c_{1}^{\prime}\exp\left(\frac{1}{2}(R_{1}(s)+R(s))\tau\right)+c_{2}^{\prime}\exp\left(\frac{1}{2}(R_{1}(s)-R(s))\tau\right),
\end{equation}
where $R_{1}(s)=\Phi_{f}(s)+\Phi_{g}(s)$ ,
$R(s)=\sqrt{(\Phi_{f}(s)-\Phi_{g}(s))^{2}+4\Theta_{g}\Theta_{f}}$
and the coefficients $c$ and $c^{\prime}$ are the initial
conditions (for the function and the derivative of  function) at
the $Q_{0}^{2}$ value. Therefore our solutions can be find
$$
f(s,\tau)=k_{1}(s,\tau)f(s,0)+k_{2}(s,\tau)\frac{\partial
f(s,\tau)}{\partial \tau}\vert_{\tau=0},
$$
and
\begin{equation}
g(s,\tau)=k_{1}(s,\tau)g(s,0)+k_{2}(s,\tau)\frac{\partial
g(s,\tau)}{\partial \tau}\vert_{\tau=0},
\end{equation}
where $ g(s,0)\equiv g(s,Q_{0}^{2})$ and $ f(s,0)\equiv
f(s,Q_{0}^{2})$ and the coefficient functions are defined by
\begin{equation}
k_{1}(s,\tau)=-\frac{1}{R(s)}\left[(R(s)+R_{1}(s))\exp\left(\frac{1}{2}R_{1}(s)\tau\right)\sinh\left(\frac{1}{2}R(s)\tau\right)-R(s)\exp\left(\frac{1}{2}(R(s)+R_{1}(s))\tau\right)\right],
\end{equation}
and
\begin{equation}
k_{2}(s,\tau)=\frac{1}{R(s)}\left[2\exp\left(\frac{1}{2}R_{1}(s)\tau\right)\sinh\left(\frac{1}{2}R(s)\tau\right)\right].
\end{equation}
In $v$-space, we introduce two kernels $\hat{K}_{1}(v,\tau)$ and
$\hat{K}_{2}(v,\tau)$ with respect to the inverse Laplace
transforms of the coefficients $ k_{1}(s,\tau)$ and
$k_{2}(s,\tau)$, as
\begin{equation}
\hat{K}_{1}(v,\tau)\equiv\mathcal{L}^{-1}
\left[k_{1}(s,\tau);v\right]\ \ , \ \
\hat{K}_{2}(v,\tau)\equiv\mathcal{L}^{-1}
\left[k_{2}(s,\tau);v\right].
\end{equation}
Also we can consider these two parameters at the initial scale
$Q^{2}=Q_{0}^{2}$ (i.e., $\tau=0$) where
$\hat{K}_{1}(v,0)=\delta(v)$ and $\hat{K}_{2}(v,0)=0$. Nothing
that the inverse Laplace transforms of the parton distributions at
the initial scale are given by
\begin{equation}
\hat{F}_{2}^{s}(v,Q^{2}_{0})=\mathcal{L}^{-1}\left[f(s,Q_{0}^{2})\right]
\ \ ,\ \
\hat{G}(v,Q^{2}_{0})=\mathcal{L}^{-1}\left[g(s,Q_{0}^{2})\right].
\end{equation}
Finally, we can obtain our general  solutions for the decoupling
of the singlet and gluon distributions with respect to Eqs.14 in
$x$-space in the following forms
 \begin{equation}
 F_{2}^{s}(x,Q^{2})=\int_{x}^{1}K_{1}\left(\frac{x}{y},\tau\right)F_{2}^{s}(y,Q^{2}_{0})\frac{dy}{y}+\int_{x}^{1}K_{2}\left(\frac{x}{y},\tau\right)\frac{\partial F_{2}^{s}(y,\tau)}{\partial \tau}\vert_{Q^{2}=Q^{2}_{0}}\frac{dy}{y},
 \end{equation}
 and
  \begin{equation}
 G(x,Q^{2})=\int_{x}^{1}K_{1}\left(\frac{x}{y},\tau\right)G(y,Q^{2}_{0})\frac{dy}{y}+\int_{x}^{1}K_{2}\left(\frac{x}{y},\tau\right)\frac{\partial G(y,\tau)}{\partial
 \tau}\vert_{Q^{2}=Q^{2}_{0}}\frac{dy}{y},
 \end{equation}
or, in general form, we can rewrite the decoupled solutions as a
reduced solution with the same coefficient functions
as
\begin{eqnarray}
\mathbb{F}(x,Q^{2})=\int_{x}^{1}K_{1}\left(\frac{x}{y},\tau\right)\mathbb{F}(y,Q^{2}_{0})\frac{dy}{y}
+\int_{x}^{1}K_{2}\left(\frac{x}{y},\tau\right)\frac{\partial
\mathbb{F}(y,\tau)}{\partial
\tau}\vert_{Q^{2}=Q^{2}_{0}}\frac{dy}{y},~~(\mathbb{F}=F_{2}^{s}~
\mathrm{or}~G).
\end{eqnarray}
To get the solution this general equation, we need to the input
initial conditions ($\mathbb{F}_{0} ~ \mathrm{and}~ \partial
\mathbb{F}_{0}$) individual for any distribution function
according to the  Refs.[11-12]. These results are  general and
gives us the exact functions  of $x$ and  $Q^{2}$ in a domain
$x_{min}{\leq}x{\leq}x_{max}$ and
$Q^{2}_{min}{\leq}Q^{2}{\leq}Q^{2}_{max}$.\\
%%%%%%%%%%%%%%%%%%%%%%%%%%%%%%%%%%%%%%%%%%%%%%%%%%%%%%%%%%%%%%
\section{Results and Conclusion}
In this method we have determined the $x$ space distribution
functions individual with respect to the initial and derivative of
initial conditions at the stating point. We have constructed a
general solution for the DGLAP evolution equations for the gluon
and singlet distributions into the initial conditions
respectively. To determine of the proton structure function we
need to know the initial singlet structure functions only and also
the initial gluon distribution functions for the gluon density
which are given in Appendix A. For the validity and compatibility
of the two sets of analytical solutions for the singlet structure
function and gluon distribution function, we compared the
distribution functions with the H1 data [14] and with the MSTW
numerical solutions [13]. We compared the solutions of Eqs.(19)
and (20) for the singlet structure function and gluon distribution
respectively with the MSTW distributions in Figs.1 and 2. Here the
input for $\{F^{s}_{2}(x,Q_{0}^{2}), \partial
F^{s}_{2}(x,Q_{0}^{2})\}$ and $\{G(x,Q_{0}^{2}), \partial
G(x,Q_{0}^{2})\}$ are taken from the Eqs.A1-A5 corresponding to
the lowest-$Q^{2}$ point. In Fig.3 we  have compared our results
for the proton structure function with  the H1 collaboration [14]
data at small $x$ at different $Q^{2}$ values. We consider the
range $45 \leq Q^{2} \leq 800~GeV^{2}$ and $0.0004< x <0.05$ for
the experimental  data. In the general solutions of singlet
structure function our computed values of $F^{s}_{2}(x,Q^{2})$
from Eq.(19) are plotted against $x$ for different values of
$Q^{2}$. In these graphs, it is observed that the best-fit curves
are obtained for $F^{p}_{2}(x,Q^{2})\simeq
\frac{5}{18}F^{s}_{2}(x,Q^{2})$ as inputs for
$\{F^{s}_{2}(x,Q_{0}^{2}), \partial F^{s}_{2}(x,Q_{0}^{2})\}$ are
taken from Eqs.A1-A3. It can be seen that our
results are in  good consistency with the H1 data.\\
In conclusion, we obtained two decoupling analytical evolution
equation, as they are two homogeneous second-order differential
equation, for the singlet  $F^{s}_{2}(x,Q^{2})$ and gluon
$G(x,Q^{2})$ distribution functions from the linear DGLAP
evolution equations by using the Laplace transform method,
respectively. These equations are general and require only a
knowledge individual $F^{s}_{2}(x,Q_{0}^{2})$,
$G_{0}(x,Q_{0}^{2})$ and those derivatives at the starting value
$Q_{0}^{2}$ for the evolution, respectively. We observed that
the general solutions are in good consistency with the experimental data and  parameterizations.\\

%\subsection{Acknowledgment}
 %G.R.Boroun is grateful to Prof.P.Ha for reading and useful comments.\\

%%%%%%%%%%%%%%%%%%%%%%%%%%%%%%%%%%%%%%%%%%%%%%%%%%%%%%
\newpage
\subsection{References}
1. Yu. L.Dokshitzer, Sov.Phys.JETPG {\bf6}, 641(1977 ).\\
2. G.Altarelli and
G.Parisi, Nucl.Phys.B{\bf126}, 298(1997 ).\\
3. V.N.Gribov and L.N.Lipatov, Sov.J.Nucl.Phys.{\bf28}, 822(1978).\\
4.M.M.Block, L.Durand, P.Ha D.W.Mckay, Phys.Rev.D\textbf{83},
054009(2011).\\
5.M.M.Block, L.Durand, P.Ha and D.W.Mckay, arXiv:1004.1440(2010).\\
6.M.M.Block, L.Durand, P.Ha D.W.Mckay, Eur.Phys.J.C\textbf{69},
425(2010).\\
7.G.R.Boroun and B.Rezaei,Eur. Phys. J. C{\bf73}, 2412(2013).\\
8.M.M. Block, Eur. Phys. J. C, {\bf65}, 1(2010).\\
9.M.M. Block, L. Durand, and D.W. McKay, Phys. Rev. D, {\bf79},
014031(2009).\\
10.M.M. Block, L. Durand, and D.W. McKay, Phys. Rev. D, {\bf77},
094003(2008).\\
11.M.M. Block, E.L. Berger and C-I. Tan, Phys. Rev. Lett. {\bf97},
252003(2006).\\
12. E.L. Berger, M.M. Block and C-I. Tan,
Phys. Rev. Lett. {\bf98}, 242001(2007).\\
13.A. D. Martin, W. J. Stirling, R. S. Thorne, and G. Watt, Eur.
Phys. J. C \textbf{63}, 189 (2009).\\
14.V. Andreev and et al. [H1 Collaboration], Eur. Phys. J. C {\bf74}, 2814(2014).\\

%%%%%%%%%%%%%%%%%%%%%%%%%%%%%%%%%%%%%%%%%%%%%%%%%%%%%%%%%%%%%%%%%%%%%%%%%%%%%%%%%%%%%%%%%%%%%%%%%%%%%%%

%%%%%%%%%%%%%%%%%%%%%%%%%%%%%%%%%%%%%%%%%%%%%%%%%%%%%%
\begin{figure}[b]
\begin{minipage}[b]{0.80\linewidth}
\includegraphics[width=1\textwidth]{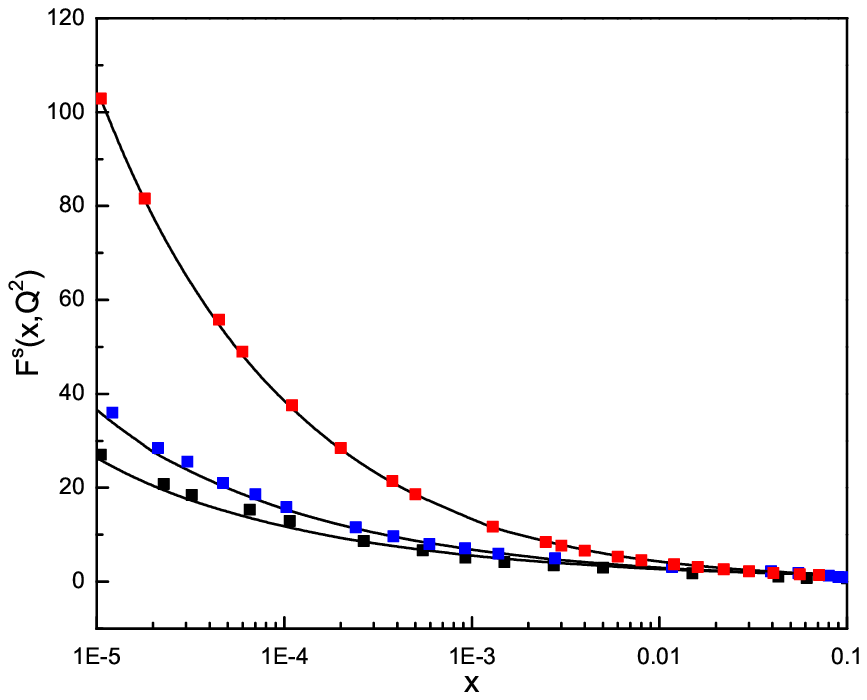}
\label{fig:minipage2}
\end{minipage}
\caption{The singlet structure function compared with the MSTW
parameterization [13](solid line) at $Q^{2}=20, 100 ~
\mathrm{and}~ M_{z}^{2}~ GeV^{2}$( bottom to top, respectively).}
\centering
\begin{minipage}[b]{0.80\linewidth}
\includegraphics[width=1\textwidth]{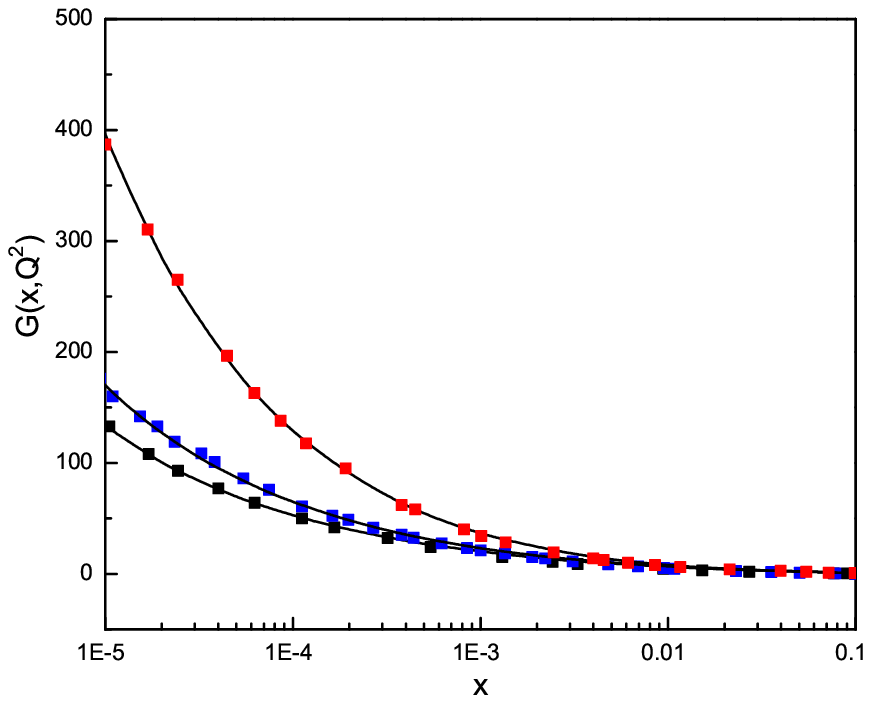}
\label{fig:minipage1}
\end{minipage}
\caption{ The same as Fig.1 for the gluon distribution function.}
\end{figure}
\begin{figure}[b]
\centering
\begin{minipage}[b]{0.80\linewidth}
\includegraphics[width=1\textwidth]{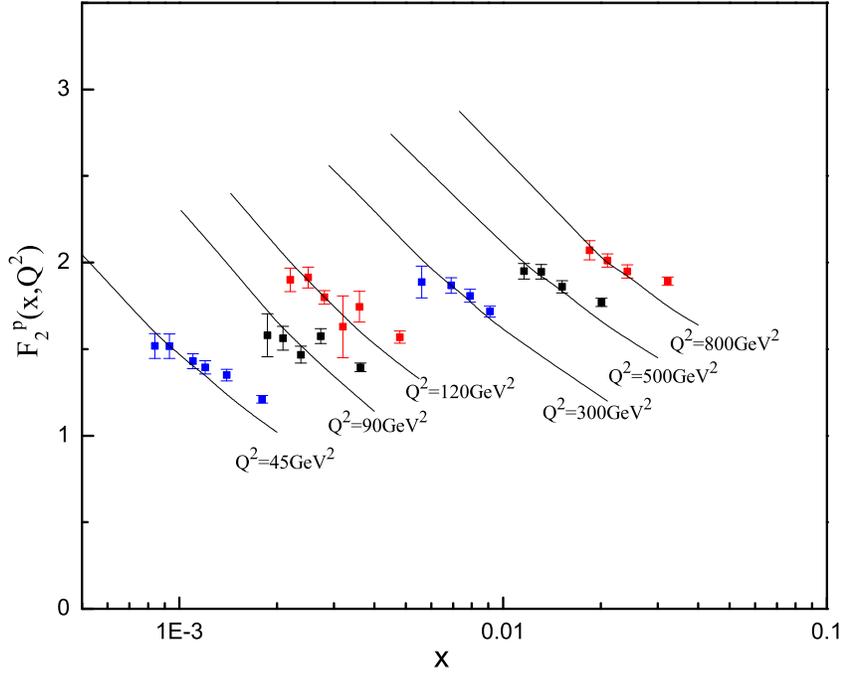}
\label{fig:minipage1}
\end{minipage}
\caption{ Proton structure function $F_{2}^{p}(x,Q^{2})$ as
function of $x$ at different $Q^{2}$ values compared with the H1
data [14].}
\end{figure}
%%%%%%%%%%%%%%%%%%%%%%%%%%%%%%%%%%%%%%%%%%%%%%%%%%%%%%%%%
\newpage
 \renewcommand{\theequation}{A-\arabic{equation}}
  % redefine the command that creates the equation no.
  \setcounter{equation}{0}  % reset counter
  \section*{Appendix A}  % use *-form to suppress numbering

In Refs.[11-12] the proton structure function $F_{2}^{p}$ for
$0<x<x_{P}$ has been parameterized in $x$-space as

$$
F_{2}^{p}(x,Q^{2})=(1-x)\Biggl(\frac{F_{P}}{1-x_{P}}+\left(a_{0}+a_{1}\ln(Q^{2})+a_{2}\ln^{2}(Q^{2})\right)\ln\left[\frac{x_{P}(1-x)}{x(1-x_{P})}\right]
$$
 \begin{align}
+
\left(b_{0}+b_{1}\ln(Q^{2})+b_{2}\ln^{2}(Q^{2})\right)\ln^{2}\left[\frac{x_{P}(1-x)}{x(1-x_{P})}\right]\Biggr),
\end{align}
where $x_{P} = 0.09$ and  $F_{2}^{p}(x_{P},Q^{2})=F_{P}=0.41$.
Table I shows fitted quantities and their errors. In region
$x<0.1$ we can approximately write $F_{2}^{s}(x,Q^{2})\simeq 18/5
F_{2}^{p}(x,Q^{2})$. So in $Q^{2}_{0}=1$ we have
\begin{equation}
\frac{\partial F_{2}^{p}}{\partial
\tau}(x,Q^{2})\vert_{Q^{2}=Q_{0}^{2}}\simeq
\frac{4\pi}{\alpha_{s}(Q_{0}^{2})}(1-x)\left[a_{1}\ln\left(\frac{x_{P}(1-x)}{x(1-x_{P})}\right)+b_{1}\ln^{2}\left(\frac{x_{P}(1-x)}{x(1-x_{P})}\right)\right],
\end{equation}
or
\begin{equation}
\frac{\partial F_{2}^{s}}{\partial
\tau}(x,Q^{2})\vert_{Q^{2}=Q_{0}^{2}}\simeq
\frac{72\pi}{5\alpha_{s}(Q_{0}^{2})}(1-x)\left[a_{1}\ln\left(\frac{x_{P}(1-x)}{x(1-x_{P})}\right)+b_{1}\ln^{2}\left(\frac{x_{P}(1-x)}{x(1-x_{P})}\right)\right].
\end{equation}
Also, the LO gluon distribution function $G(x,Q^{2}) =
xg(x,Q^{2})$ has been determined by
$$
G(x,Q^{2})=c_{0}+c_{1}\ln(Q^{2})+c_{2}\ln^{2}(Q^{2})+\left[d_{0}+d_{1}\ln(Q^{2})+d_{2}\ln^{2}(Q^{2})\right]\ln\left(\frac{1}{x}\right)
$$
\begin{equation}
\left[e_{0}+e_{1}\ln(Q^{2})+e_{2}\ln^{2}(Q^{2})\right]\ln^{2}\left(\frac{1}{x}\right).
\end{equation}
 The parameters have been shown in Table I. The derivative of LO gluon distribution function in $Q^{2}_{0}=1$ becomes
\begin{equation}
\frac{\partial G}{\partial
\tau}(x,Q^{2})\vert_{Q^{2}=Q_{0}^{2}}=\frac{4\pi}{\alpha_{s}(Q^{2}_{0})}\left(c_{1}+d_{1}\ln\left(\frac{1}{x}\right)+e_{1}\ln^{2}\left(\frac{1}{x}\right)\right).
\end{equation}
%%%%%%%%%%%%%%%%%%%%%%%%%%%%%%%%%%%%%%%%%%%%%%%%%%%%%%%%%%%%%%%%%%%%%%%%%%%%%%%%%%%%%%%%%%%%%%%%%%%%%%%%%%%%%%%%%%%%%%%%%%%%%%

\begin{table}[h]
\caption{ The fitted parameters}
\begin{tabular} {cccc}
\toprule \\  \multicolumn{2}{c}{parameters \quad \quad  \quad \quad value}  \quad \quad \quad \quad  & \multicolumn{2}{c}{parameters\quad  \quad \quad \quad \quad \quad \quad \quad value } \\ &&&\\ \hline \\ &&&\\
$c_{0}$ &  \quad \quad \quad \quad $-0.459$ & \quad \quad \quad \quad  $a_{0} $  &  \quad \quad \quad \quad  $-5.381\times 10^{-2}\pm2.17\times10^{-3} $  \\
 $c_{1}$ &  \quad \quad \quad \quad $-0.143$ & \quad \quad \quad \quad $a_{1} $  &  \quad \quad \quad \quad   $2.034\times 10^{-2}\pm1.19\times10^{-3}$  \\
$c_{2}$ & \quad \quad \quad \quad $ -0.0155$  & \quad \quad \quad \quad  $a_{2}$   &   \quad \quad \quad \quad  $4.999\times 10^{-3}\pm2.23\times10^{-4}$   \\  &&&\\
$d_{0}$&  \quad \quad \quad \quad $0.231$ &  \quad \quad \quad \quad $b_{0}$   &   \quad \quad \quad \quad  $9.955\times 10^{-3}\pm3.09\times10^{-4} $ \\

$d_{1}$ &  \quad \quad \quad \quad $0.00971$ &  \quad \quad \quad \quad $b_{1}$   &   \quad \quad \quad \quad  $3.810\times 10^{-3}\pm1.73\times10^{-4}$  \\

$d_{2}$ & \quad \quad \quad \quad $-0.0147$ & \quad \quad \quad \quad $b_{2}$    &    \quad \quad \quad \quad $9.923\times 10^{-4}\pm2.85\times10^{-5} $ \\ &&& \\
$e_{0}$& \quad \quad \quad \quad $0.0836$ & &\\
$e_{1}$ & \quad \quad \quad \quad  $0.06328$ & &\\
$e_{2}$ &  \quad \quad \quad \quad $0.0112$ & &\\ \hline

\end{tabular}
\end{table}

\end{document}